\documentclass{article}

\usepackage{arxiv}

\usepackage[utf8]{inputenc} 
\usepackage[T1]{fontenc}    
\usepackage{hyperref}       
\usepackage{url}            
\usepackage{booktabs}       
\usepackage{amsfonts}       
\usepackage{nicefrac}       
\usepackage{microtype}      
\usepackage{lipsum}		
\usepackage{graphicx}
\usepackage{natbib}
\usepackage{doi}

\usepackage{amsmath}
\usepackage{graphicx}
\usepackage{amssymb}
\usepackage{booktabs}
\usepackage{color}
\usepackage{epsfig,amsthm}
\usepackage{hyperref} 
\usepackage{subcaption}

\title{Optimal Execution Using Reinforcement Learning}

\author{ Cong ZHENG\\
        \texttt{czhengae@connect.ust.hk}\\
 Department of Mathematics\\
 The Hong Kong University of Science and Technology\\
        \And Jiafa He \\
        \texttt{jhebu@connect.ust.hk}\\
        Department of Mathematics\\
 The Hong Kong University of Science and Technology\\
        \And  Can YANG\\
        \texttt{macyang@ust.hk}\\
 Department of Mathematics\\
 The Hong Kong University of Science and Technology\\
}

\renewcommand{\shorttitle}{\textit{arXiv} Template}

\hypersetup{
pdftitle={A template for the arxiv style},
pdfsubject={q-bio.NC, q-bio.QM},
pdfauthor={David S.~Hippocampus, Elias D.~Striatum},
pdfkeywords={First keyword, Second keyword, More},
}

\begin{document}
\maketitle

\begin{abstract}
	This work is about optimal order execution, where a large order is split into several small orders to maximize the implementation shortfall. Based on the diversity of cryptocurrency exchanges, we attempt to extract cross-exchange signals by aligning data from multiple exchanges for the first time. Unlike most previous studies that focused on using single-exchange information, we discuss the impact of cross-exchange signals on the agent's decision-making in the optimal execution problem. Experimental results show that cross-exchange signals can provide additional information for the optimal execution of cryptocurrency to facilitate the optimal execution process.
\end{abstract}

\keywords{optimal execution, cross-exchange, reinforcement learning}

\section{Introduction}

\subsection{Background}
Optimal execution is a fundamental problem in the financial field, first proposed by Bertsimas and Lo in 1998\cite{bertsimas1998optimal}. This problem addresses how traders can buy or sell a given quantity of stocks within a certain time frame while minimizing trading costs. The background of this problem is the block trading in modern securities markets. Institutional investors or high-net-worth individuals often need to engage in large-scale asset allocation, but the market may not be able to provide the corresponding liquidity in the short term. Therefore, block trades have to be split into a large number of small orders to minimize market impact while minimizing trading costs. With the emergence of electronic trading platforms in the late 20th century, algorithmic trading via computers began to emerge. Traders have started to design execution algorithms that consider multiple factors such as market liquidity, trading fees, market sentiment, and more, hoping to improve trading efficiency while reducing trading costs. These algorithms use mathematical models to optimize execution strategies, considering various constraints and objectives, such as minimizing market impact, managing risk exposure, and achieving better execution prices.

Another important background to consider is that most trading strategies are highly sensitive to trading costs. A commonly used metric for measuring the profitability of a trading strategy is ROT (return on turnover), which is typically measured in basis points (bps), with one bps representing 0.01\%. ROT represents the profitability of a unit of trading volume, and if the ROT of a strategy can cover the trading costs, the strategy can be profitable. Trading costs consist of two parts: transaction fees charged by the exchange and the wear and tear of the trading process. Edelen et al. (2013)\cite{edelen2013shedding} conducted a study on the "invisible" costs of fund trading and found that a significant portion of the trading costs comes from the impact on the stock price when buying and selling stocks. The process of buying and selling stocks usually causes a reverse fluctuation in the stock price, which increases trading costs. Therefore, reducing trading costs is essential to improving the profitability of a strategy.

In recent years, electronic trading platforms such as NASDAQ have become increasingly prevalent, with trades on both the buyer and seller sides typically conducted through an order book. There are two main categories of order execution methods: limit order execution and market order execution. For example, when we need to buy a certain amount of stock, we can wait for the order to be filled by placing a limit buy order, or we can fill the order directly by taking the sell order on the sell side. Limit order execution, usually placed in the form of a limit order, is considered a passive order placement strategy. Its characteristic is that there are fewer opportunities for execution, but the execution price is usually better. Market order execution, on the other hand, is usually placed in the form of a market order. Its characteristic is that the execution speed is faster, but the execution price is usually worse due to the crossing of the bid-ask spread. Market orders are typically used to avoid missing out on opportunities, while limit orders are used to achieve better execution prices. Optimal execution aims to balance the trade-off between execution speed and execution price.

The time-weighted average price (TWAP) strategy and volume-weighted average price (VWAP) strategy provide a solid baseline for optimal execution problems. These two execution algorithms do not require any assumptions or predictions about the market and only focus on buying or selling a given quantity of stocks at a specified time while minimizing the impact on the market. TWAP executes trades evenly over a specified time period to achieve an average execution price that is close to the time-weighted average price over that period. This approach is effective in minimizing market impact, but it may not be suitable for highly volatile markets or when market conditions change rapidly.

To optimize order transactions, it is crucial to consider market conditions and adopt different order execution strategies in different market environments. However, integrating our own inventory status with market conditions presents a significant challenge to theoretical modeling. Reinforcement learning models offer a promising solution. An intelligent agent can make decisions by considering its own inventory and market conditions, and learn decision-making experience by interacting with the market and receiving feedback.

\subsection{Related Work}
Traditionally, the optimal execution problem has been modeled as a stochastic control problem. Bertsimas and Lo (1998)\cite{bertsimas1998optimal} used dynamic programming to model stock prices and price impact functions, and provided a closed-form solution for optimal execution. Under their assumptions, the transaction can be completed within a fixed time frame while minimizing the transaction execution costs. Another early work is the AC model (Almgren and Chriss, 2001)\cite{almgren2001optimal}, which provides an optimal solution for optimal execution under the basic assumption that stock prices follow Brownian motion. Similar work includes Huberman and Stanzl (2005)\cite{huberman2005optimal}, which also models optimal execution as a stochastic control problem and provides an analytical solution by minimizing implementation shortfall. While closed-form solutions provide some intuitive explanations for the optimal execution problem, they make strong assumptions about the distribution of underlying asset price changes, which may not be fully consistent with the distribution of actual price data. In the complex background of real applications, it is difficult to provide theoretical solutions that fully account for all market conditions and trading strategies.

As financial markets continue to evolve, the amount of financial data available is growing rapidly, with various trading platforms and markets providing vast amounts of trading data. In recent years, researchers have started using real price data instead of simulated price data, leading to significant advances in optimal execution algorithms. Nevmyvaka et al. (2006)\cite{nevmyvaka2006reinforcement} presented the first large-scale application of data-driven optimal execution algorithms. They used a reinforcement learning algorithm, Q-learning, to solve the optimal execution problem. Their experimental results showed that by considering some market states, reinforcement learning can significantly reduce execution costs compared to traditional execution algorithms. With the continued development of reinforcement learning, more advanced models have been developed to solve the optimal execution problem. Dahlen et al. (2018) and Schnaubelt et al. (2022) \cite{dahlen2018optimized, schnaubelt2022deep} have used DDQN and PPO algorithms, respectively, to solve the optimal execution problem. These algorithms have overcome the shortcomings of Q-learning in solving high-dimensional problems.

In addition to determining the optimal placement price, another important area of research in algorithmic trading is order scheduling, which refers to how to split large orders into smaller market orders. Unlike TWAP's uniform execution, order scheduling requires making more nuanced decisions based on the current market environment. There has been a significant amount of research on order scheduling, with many studies using reinforcement learning algorithms to optimize this process. Ning et al. (2021)\cite{ning2021double} used deep double Q-networks to implement order scheduling and demonstrated superior performance compared to traditional scheduling algorithms on nine different stocks. Other studies on using reinforcement learning to solve order scheduling include Bao et al. (2019)\cite{bao2019multi}.

Using reinforcement learning to solve the optimal execution problem presents two main challenges. First, it is necessary to plan the execution to complete the optimal execution task within the specified time frame. This requires making decisions on the order type, quantity, and price while taking into account market conditions such as liquidity, volatility, and trading fees. Second, it is necessary to predict short-term market trends by analyzing market conditions, including order flow, order book structure, and other information, to complete the execution task at a better price. This requires analyzing and predicting asset time series, typically focusing on predicting asset price changes in seconds or even milliseconds. To address these challenges, the reinforcement learning model needs to consider both aspects of information to complete the order execution task within the specified time.

\subsection{Our Work And Contribution}

Our work focuses on the cryptocurrency market, which has experienced explosive growth in recent years and currently has a market value of trillions of US dollars. Cryptocurrencies are encrypted electronic currencies based on cryptography, featuring decentralization and high transparency. Virtual currency trading shares similarities with traditional finance, but also exhibits distinct characteristics. One of the main distinctive features of cryptocurrency trading is that it is at an early stage of development, leading to greater price volatility compared to traditional finance. Additionally, there is a diversity of trading venues in virtual currency trading, particularly with the development of DeFi (Decentralized Finance). A significant amount of bitcoin trading volume occurs on decentralized exchanges, while there are also several equally competitive centralized exchanges. The diversity of trading venues presents challenges for analyzing asset time series, as asset prices are not only influenced by transactions on a single exchange but also by the price movements on all exchanges. Trading on DeFi platforms also has a significant impact on asset prices. When trading on a single exchange, analyzing time series requires considering not only the current transaction info from that exchange but also information from other exchanges to obtain a comprehensive understanding of market conditions.

Our work focuses on studying optimal order execution in the cryptocurrency market and includes several key aspects. First, we collected virtual currency data from several mainstream exchanges and conducted an extensive analysis of the process of asset price formation. This analysis included studying how asset prices change and how price signals are transmitted between exchanges, to better understand the market environment in which we are operating. Second, we established a virtual exchange environment to simulate the order-matching process. In this environment, agents can obtain feedback from the virtual exchange by providing order execution decisions, including order type, order direction, order price, and order quantity. Thirdly, by incorporating cross-exchange price signals, we can use advanced reinforcement learning algorithms to train agents to better complete order execution tasks. This is also the first attempt to study optimal order execution using cross-exchange signals. Compared to the traditional TWAP algorithm, the reinforcement learning model can further reduce trading friction, leading to better execution outcomes. In addition to the aforementioned aspects, we also investigate the impact of the prediction interval on the optimal execution problem. By analyzing the agent's action space, we can also discover how cross-exchange signals affect the agent's decision-making. This can provide insights into how to better incorporate cross-exchange information into execution strategies and improve overall execution performance.

In Section 3.2, we mainly discuss the collection and preprocessing of cross-exchange data. We focus on the collection process of high-frequency data and the data processing process, which includes data cleaning, normalization, and feature engineering. Section 3.3 introduces the micro-structure of high-frequency data and how it can be used to predict future asset price fluctuations. We also discuss the application of cross-market signals and how they can be incorporated into prediction models. Section 3.4 is devoted to modeling optimal execution processes. We discuss how to model reinforcement learning states, actions, and reward functions, and how to incorporate cross-exchange signals into these models. Finally, we present experimental results and compare our approach with traditional methods. Our experiments demonstrate that our approach outperforms traditional methods in terms of execution quality and trading costs, and highlight the importance of incorporating cross-exchange signals in optimal execution strategies.

\section{High-Frequency Data And Preprocess}
Our work focuses on analyzing perpetual data from several top-ranked cryptocurrency exchanges, including Binance, OKEx, and Bybit. Perpetuals are a type of derivative that exhibit similar price fluctuations to spot prices but can be traded with higher leverage. Perpetuals typically have higher trading volumes than spot markets. It is important to note that the prices of spot and perpetual markets are interrelated and balanced through funding rates. When the perpetual price is higher than the spot price, perpetual longs need to pay a certain fee to shorts regularly, and when the perpetual price is lower than the spot price, perpetual shorts need to pay a certain fee to longs regularly to maintain the spot-perpetual price difference within a certain range. Given that perpetual trading markets have more active transactions than spot trading markets, we focus on analyzing the perpetual price data from these three exchanges. Our analysis includes studying the micro-structure of the data and identifying the key factors that influence price movements in the perpetual markets.

Most mainstream virtual currency exchanges now provide interfaces for downloading high-frequency historical data. While it is possible to analyze the time series information of a single exchange by downloading the corresponding historical data, analyzing and comparing the price time series of multiple exchanges requires additional processing. The process of horizontally analyzing price data from multiple exchanges requires aligning the timestamps of the price data from each exchange. However, the historical data provided by the exchanges usually only contains their own timestamps, making timestamp alignment difficult to achieve. To address this issue, we use WebSockets to centrally collect and process data from multiple exchanges in real-time and align the data from the different exchanges using local timestamps. Websockets are a new protocol based on HTML5 that enables full-duplex communication between the user side and the server side, allowing data to be quickly transmitted in both directions. By using WebSockets, we can collect high-frequency data from multiple exchanges simultaneously and process it in real-time. This allows us to align the timestamps from the different exchanges and perform a horizontal analysis of the price data.

The data we collect can be divided into two categories: multi-level order book data and real-time market transaction data. Transaction data is updated in real-time and typically includes four pieces of information: transaction quantity, transaction price, transaction direction, and transaction time. Order book data, on the other hand, can have different update rules across exchanges. For example, when subscribing to a 50-level order book on OKEx, the order book update frequency is usually around 10ms, while on Binance, the order book is typically updated every 100ms for the incremental part. Order book data can be divided into two parts: seller's order book information and buyer's order book information. Order book information includes two parts: the price of each level and the corresponding order quantity. In addition to subscribing to the order book data of each exchange, we also simultaneously subscribe to the best bid-ask data to obtain more real-time data. We have noticed that the best bid-ask data usually has a higher update frequency than the order book, which is crucial for high-frequency price prediction. For example, on Binance, the order book data is only updated every 100ms, which may not meet our needs during periods of large market movements. During these 100ms, the order book may have changed significantly and the best bid-ask data may have been updated several times. To address this issue, we locally maintain a relatively real-time order book by combining ticker data with exchange order book data. This allows us to obtain more real-time data and improve the accuracy of our price prediction models.

\begin{figure}
	\begin{center}
	\includegraphics[width=0.98\columnwidth]{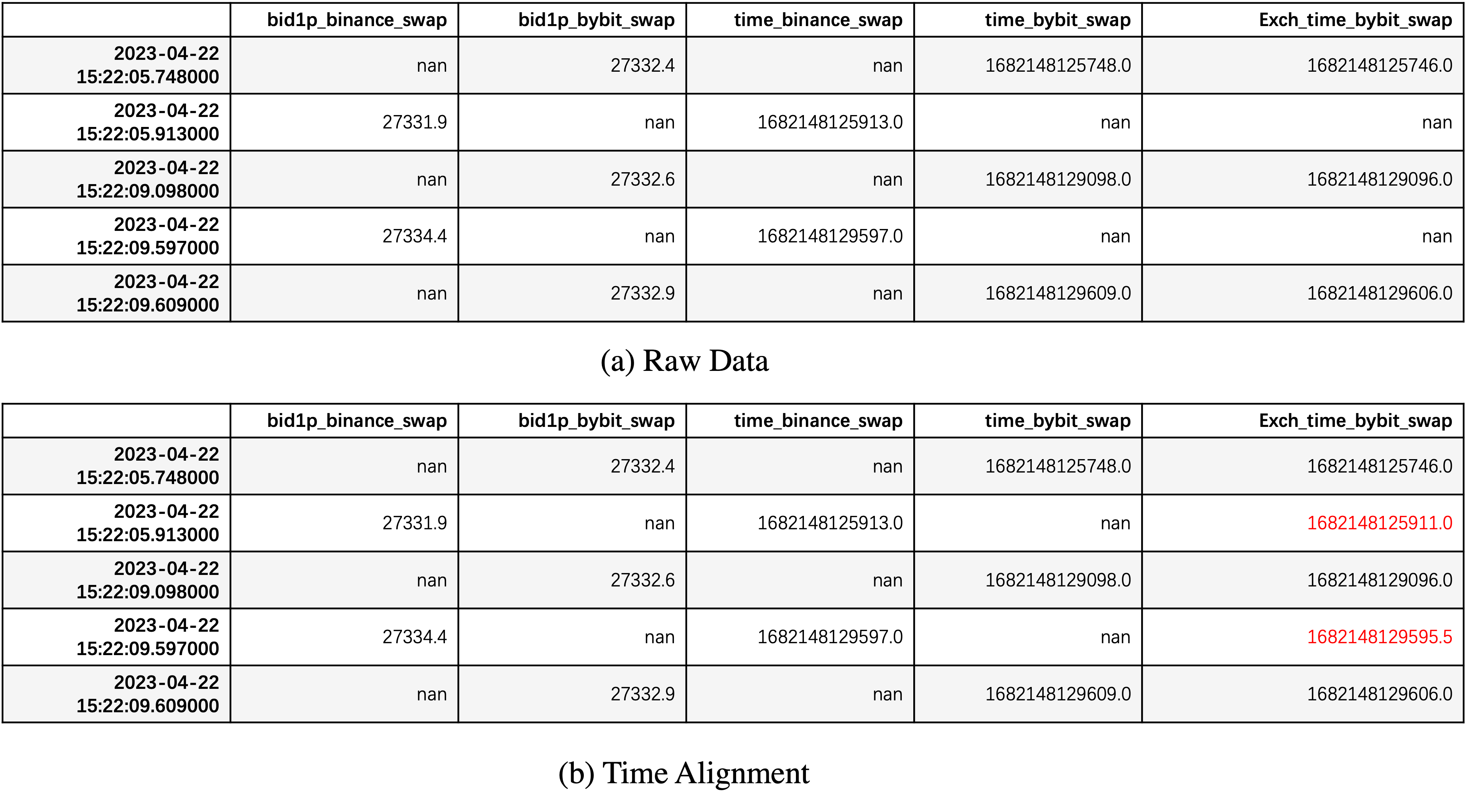}
	\end{center}
	\caption{\textbf{Time Alignment Example.} By using a single server to collect data from different exchanges, we can obtain the server time ($time\_binance\_swap$, $time\_bybit\_swap$) for each piece of data received. For the exchange Bybit, we also record the corresponding exchange time ($Exch\_time\_bybit\_swap$). The time alignment process involves mapping the server time to the exchange time for all exchanges using an interpolation method.}
	\label{Time alignment example}
\end{figure}

Another challenge we face when collecting data from multiple exchanges is that the servers of different exchanges are located at different addresses. For example, OKEx's server is located in Hong Kong, Binance's server is located in Tokyo, and Bybit's server is located in Singapore. This presents a challenge for aligning the timestamps of different exchanges. Traditionally, high-frequency traders choose trading servers that are closest to the target exchange to obtain trading data as quickly as possible, as shortening the physical distance is the most direct way to reduce network latency. However, when we need data from multiple exchanges simultaneously, we inevitably face the problem of larger latency due to the larger physical distance. To address this challenge, we use a server in Hong Kong to collect data from various exchanges. Although there is inevitably some delay in data from Binance or Bybit due to the larger physical distance, this solution solves the problem of aligning timestamps and avoids forward-looking bias. The time alignment process can be seen in figure \ref{Time alignment example}.

For data preprocessing, we used a fixed time interval sampling method with a fixed 10ms sampling interval. The high-frequency data collected is event-driven, and the time intervals are usually not fixed or irregular. However, using a fixed 10ms sampling interval can greatly simplify the analysis of factors. The fixed time interval sampling method also brings great convenience to the horizontal comparison of exchange price data and the extraction of cross-exchange signals. By aligning the data from multiple exchanges using the local timestamps, we can perform a horizontal analysis of the price data and extract cross-exchange signals. In addition, the fixed time interval sampling method also enables us to perform more efficient and effective data analysis. By using a fixed time interval, we can easily aggregate and summarize the data, making it easier to identify patterns and trends in the data.

\section{Cross Exchange Feature Engineer}
Unlike the traditional financial system, where prices are typically driven by a single exchange, the trading venues for virtual currencies are more diverse, and the reasons for asset price changes are more varied. To understand the causes of exchange rate fluctuations, we need to collect and analyze information from multiple exchanges. This section focuses on the relationship between cross-exchange signals and short-term asset price changes. These signals are mainly extracted from the market order flow, which drives asset price changes. Limit order flow includes placing and canceling limit orders, which are usually reflected in changes in the order book shape. Market order flow refers to transaction information in the market, which reflects the strength of the trader's desire to buy or sell. By analyzing the order flow data from multiple exchanges, we can extract cross-exchange signals that can help us predict short-term asset price changes. These signals can provide valuable information about the market's overall sentiment and can be used to develop more accurate and effective models of the cryptocurrency market.

\subsection{Cross exchange Trade Imbalance}
Market orders are often considered the main driver behind asset price movements in the cryptocurrency market. When there are more active buyers than sellers, asset prices tend to rise, while when there are more sellers than buyers, asset prices usually fall. This can be explained from the perspective of supply and demand balance. When demand exceeds supply, i.e., when there are more buyers than sellers, prices generally rise, while when supply exceeds demand, i.e., when there are more sellers than buyers, prices generally fall.

Order flow imbalance has been widely studied in the context of the cryptocurrency market, and researchers like Cont \cite{cont2021price} have conducted research on multi-level order flow imbalance. Order imbalance is typically defined as shown in Equation \ref{order flow imbalance}, where $OIM_t$ represents the statistical results of the order flow from time $t-1$ to $t$. To consider the real-time statistical results of order flow from different exchanges, we use $i$ to represent different exchanges. The "Buy volume" and "Sell volume" are the active trading volumes of traders, with "Buy volume" representing the active buying volume and "Sell volume" representing the active selling volume. Cartea \cite{cartea2016incorporating} also pointed out that order flow is the main reason for short-term price changes in the cryptocurrency market. By analyzing the dynamics of order flow and order flow imbalance, we can gain valuable insights into the market's overall sentiment and predict short-term price changes more accurately.

\begin{equation}\label{order flow imbalance}
    OIM_{t}^{i} = Buy\_volume^{i}_{t}-Sell\_volume^{i}_{t},
\end{equation}

the symbol $OIM_{t}^{i}$ represents the strength of buying and selling, where a value greater than zero indicates stronger buying power than selling power, and the absolute value represents the specific magnitude. However, there is a normalization issue that needs to be addressed. Firstly, there is a significant difference in trading volume between different assets, and secondly, even for the same asset, the trading activity may vary significantly across different trading periods. To address this normalization issue, we provide a normalized definition of order flow imbalance in Equation \ref{normalized order flow imbalance}. By normalizing the order flow imbalance using the total trading volume, we can compare the order flow imbalance of different assets or different trading periods on a more equal basis.

\begin{equation} \label{normalized order flow imbalance}
OIMN_{t}^{i} = \frac{\text{sign}(OIM_{t}^{i})\cdot(\lvert OIM_{t}^{i}\rvert - \min(\lvert OIM_{t}^{i}\rvert))}{(\max(\lvert OIM_{t}^{i}\rvert) - \min(\lvert OIM_{t}^{i}\rvert))}.
\end{equation}

In the normalized definition of order flow imbalance (Equation \ref{normalized order flow imbalance}), the symbols $Min(OIM_{t}^{i})$ and $Max(OIM_{t}^{i})$ represent the minimum and maximum values of the indicator $OIM_{t}^{i}$ over a certain period of time. The specific magnitude of the order flow imbalance is characterized by the size of the current order flow relative to the maximum and minimum values.

The R score, also known as the coefficient of determination, is a statistical measure widely used in regression analysis to represent the explanatory power of a model for the sample. If we use $y$ to represent the true future return of an asset, and $\hat{y}$ as the predicted value of the regression model, then the R score can be expressed as the following formula \ref{R score}: 

\begin{equation}\label{R score}
    R^{2} = 1- \frac{\sum(y-\hat{y})^2}{\sum(y-\Bar{y})^2}.
\end{equation}

As we can see, the value of the R score usually falls between 0 and 1. The closer it is to 1, the better the model fits the data, while the closer it is to 0, the poorer the model's explanatory power.

\begin{figure}
	\begin{center}
	\includegraphics[width=0.98\columnwidth]{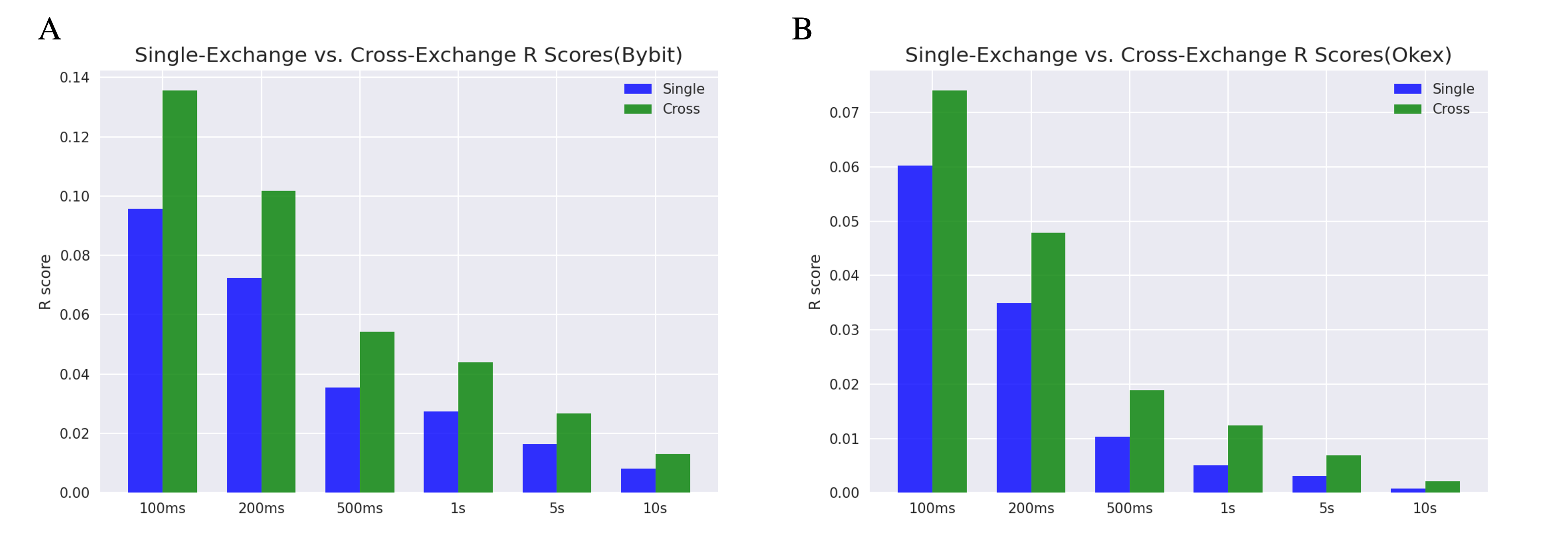}
	\end{center}
	\caption{\textbf{R score comparison (BTC\_USDT).} This figure shows the prediction performance of the trade flow of two exchanges, Bybit and Okex, respectively.}
	\label{Trade flow R2}
\end{figure}

In order to measure the predictive power of the factor, we consider the $R^{2}$ value of the linear model, as shown in Figure \ref{Trade flow R2}.

Figure A and Figure B in \ref{Trade flow R2} show the prediction performance of the trade flow of two exchanges, Bybit and Okex, respectively. Taking Figure A as an example, the x-axis represents multiple time dimensions, and we focus on the prediction performance of 100ms, 200ms, 500ms, 1s, 5s, and 10s. The y-axis shows the R score of the regression model, which measures the quality of the model's fit to the data. We also compared the effects of using signals from a single exchange and considering signals from multiple exchanges at the same time. The multi-exchange signal is derived from the single-exchange signal, as shown in Equation \ref{normalized cross exchange order flow imbalance}. For example, when we predict the future return of Bybit, we will consider the order flow situation of Binance and Okex comprehensively; when we predict the future return of OKEx, we will consider the order flow situation of Binance and Bybit comprehensively from time $t-1$ to $t$.

\begin{equation}\label{normalized cross exchange order flow imbalance}
    OIMN_{t} = \sum_{i} {OIMN_{t}^{i}}.
\end{equation}

In Figure A and Figure B, we observe that the cross-exchange signal consistently outperforms the single-exchange signal in terms of predictive power, regardless of the prediction time scale. This finding underscores the importance of incorporating cross-exchange order flow signals into predictive models of asset returns in the cryptocurrency market. We also observe that the R score of the regression model tends to decline as the prediction time length increases. This is consistent with intuition, as short-term price prediction is usually easier than long-term price prediction. However, even for longer prediction time scales, the cross-exchange signal still provides a significant improvement in predictive power compared to the single-exchange signal. Another interesting finding is that predicting the asset price of OKEx is more difficult than that of Bybit. This phenomenon can be explained from the perspective of trading volume. Binance is generally considered the largest cryptocurrency trading market with the highest trading volume, followed by Okex and Bybit. The trading volume of Okex is usually higher than that of Bybit. This difference in trading volume may contribute to the greater prediction difficulty for Okex compared to Bybit.

\begin{figure}
	\begin{center}
	\includegraphics[width=0.98\columnwidth]{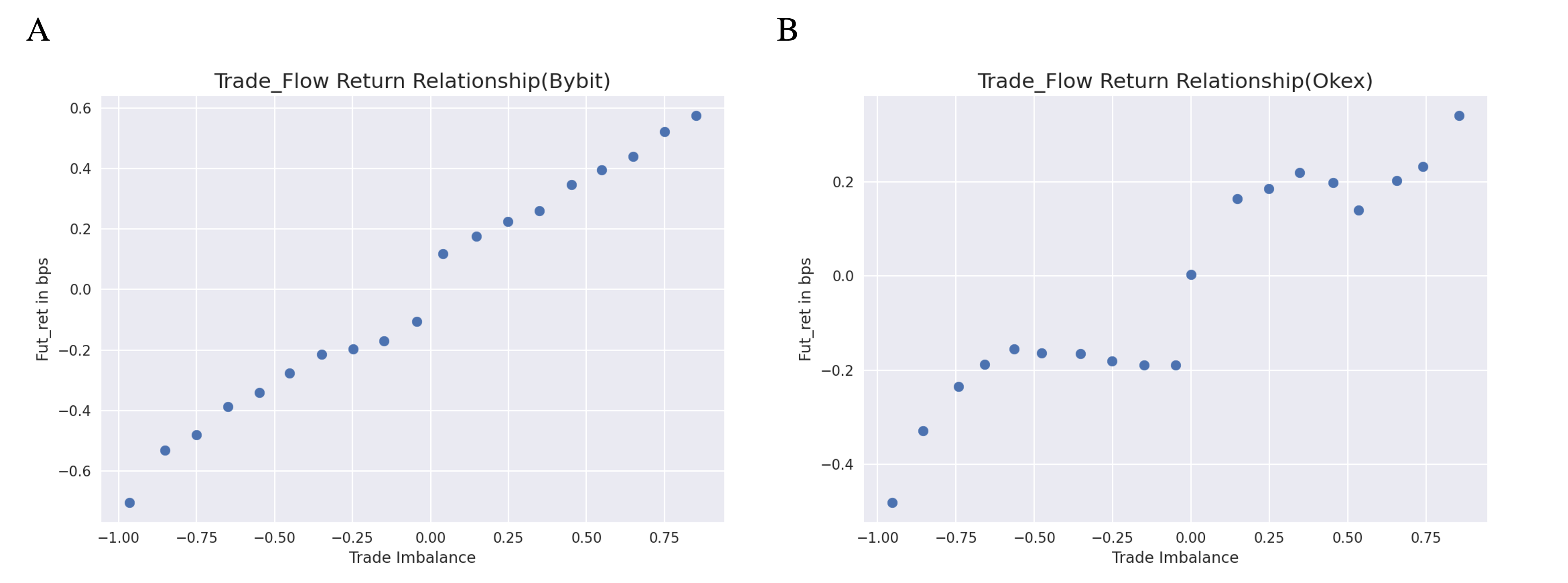}
	\end{center}
	\caption{\textbf{Future return prediction 5s (BTC\_USDT).} This figure presents the relationship curve between cross-exchange trade imbalance and future return.}
	\label{Trade flow fut return}
\end{figure}

To further analyze the relationship between trade imbalance and asset future return, we present the relationship curve between cross-exchange trade imbalance and future return in Figure \ref{Trade flow fut return}. Figure A reflects the prediction effect for Bybit, while Figure B reflects the prediction effect for OKEx. The curves demonstrate that the trade imbalance and future return roughly show a proportional relationship, and the prediction effect of Bybit is much better than that of OKEx. Taking Bybit as an example, when the trade imbalance tends to 1, the future return tends to 0.6 basis points (bps), which means that the trade imbalance can provide approximately 0.6 bps of prediction information. This finding supports the use of trade imbalance information as a predictor of short-term asset returns.

\subsection{Cross Exchange Order Imbalance}

\begin{figure}
	\begin{center}
	\includegraphics[width=0.7\columnwidth]{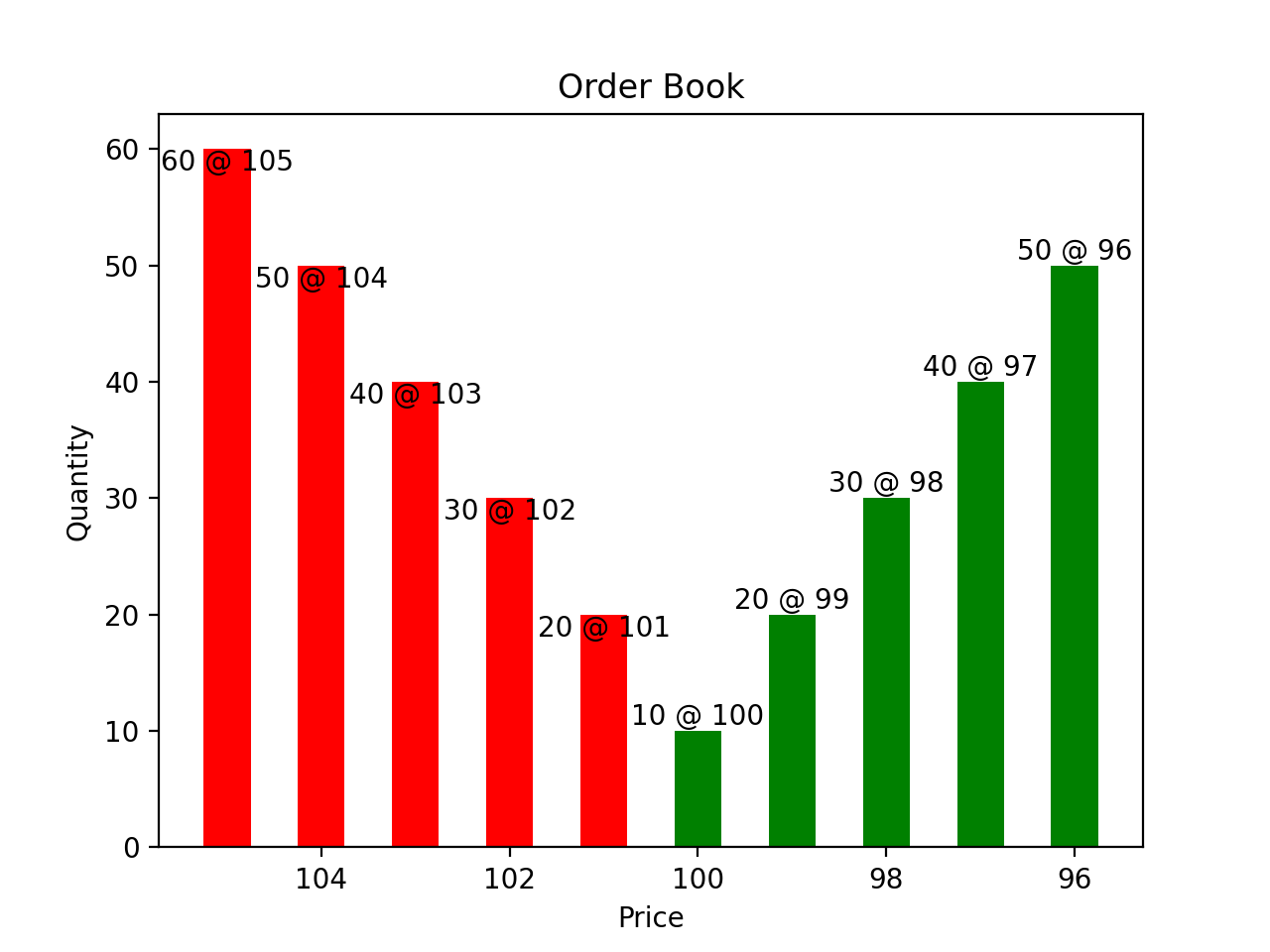}
	\end{center}
	\caption{\textbf{Order book Example} An example of a limit order book, where the horizontal axis represents the order price and the vertical axis represents the order quantity. The green part represents limit buy orders, while the red part represents limit sell orders.}
	\label{Orderbook example}
\end{figure}

High-frequency traders need to pay attention to and study the microstructure of the order book, which plays a critical role in modern financial markets. Most financial markets, including the Hong Kong, New York, Shenzhen, Tokyo, and NASDAQ exchanges, use limit order books to match trades. Unlike the previous quote-driven trading model, the limit order book is organized by all market participants, who can issue or cancel orders reflecting their desire to buy or sell assets at specific prices. Figure \ref{Orderbook example} shows an example of a limit order book, where the horizontal axis represents the order price and the vertical axis represents the order quantity. The green part represents limit buy orders, while the red part represents limit sell orders. Traders often try to obtain information about asset price changes by observing the shape or changes in the order book. For example, in the example shown in Figure \ref{Orderbook example}, the quantity of red sell orders is slightly larger than that of green buy orders. This is often interpreted as indicating that the selling pressure is greater than the buying pressure, and the probability of a price drop is higher. Traders can also analyze the dynamics of the order book, such as the rate of order cancellations and updates, to infer market sentiment and anticipate price movements.

Order book imbalance has been the subject of extensive study and analysis \cite{cont2014price, bouchaud2009markets, byrd2020importance, yagi2023impact}. Cartea \cite{cartea2018enhancing} has demonstrated that order book imbalance can be used to predict the direction of the next market order, whether it is a buy or a sell, thus predicting short-term price changes in assets. Although limit orders reflect buying and selling desires relatively weakly compared to trade flow, they can significantly impact short-term price trends by providing liquidity. In our analysis, we consider the cumulative order volume of the top five bid and ask levels and define the order book imbalance as shown in Equation \ref{orderbook imbalance}, where $i$ represents the exchange and $t$ represents the instantaneous order book imbalance at time $t$. We also provide the formula for calculating cross-exchange order book imbalance in Equation \ref{cross exchange orderbook imbalance}. Cross-exchange order book imbalance refers to predicting short-term price trends of assets by considering the shapes of order books from multiple exchanges. Based on the formula, when the cumulative liquidity of the buy side is higher than that of the sell side, $IMB_t^i$ is positive, indicating that the probability of a short-term increase in asset prices is greater than a decrease. Conversely, when the cumulative liquidity of the sell side is higher than that of the buy side, $IMB_t^i$ is negative, indicating that the probability of a short-term decrease in asset prices is greater than an increase.

\begin{equation}\label{orderbook imbalance}
    IMB_{t}^{i} = (Bid_{t}^{i}-Ask_{t}^{i})/(Bid_{t}^{i}+Ask_{t}^{i}).
\end{equation}
\begin{equation}\label{cross exchange orderbook imbalance}
    IMB_{t} = \sum_{i} {IMB_{t}^{i}}.
\end{equation}

\begin{figure}
	\begin{center}
	\includegraphics[width=0.98\columnwidth]{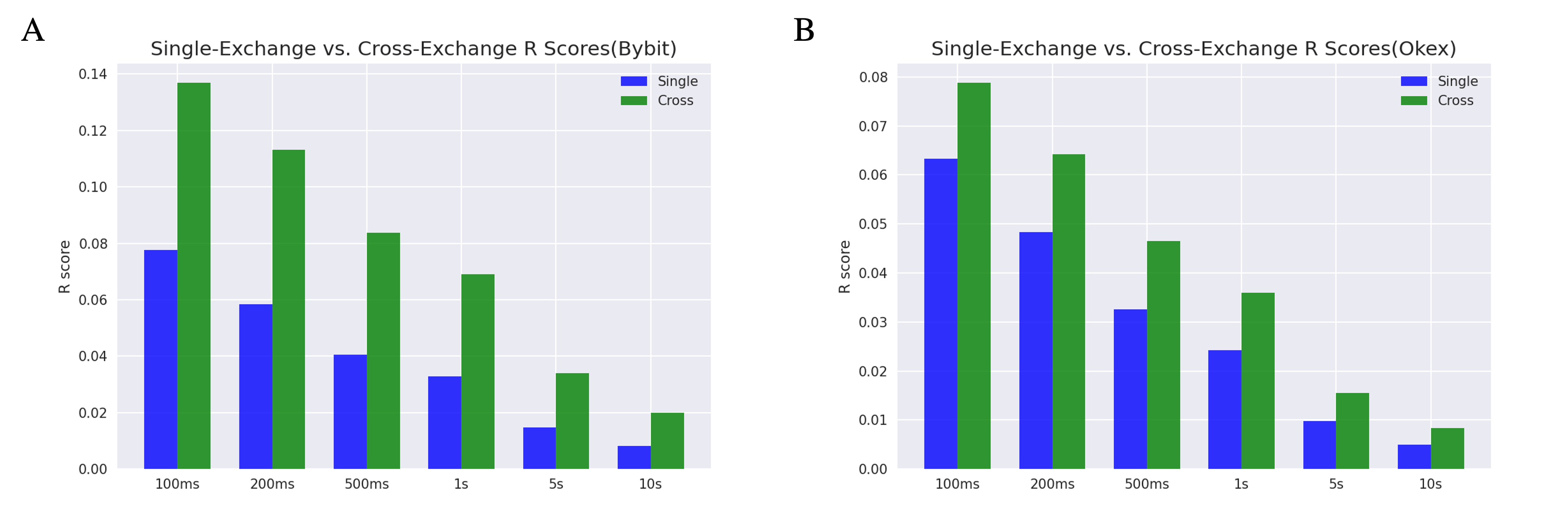}
	\end{center}
	\caption{\textbf{R Score Comparison (BTC\_USDT).} This figure demonstrates the predictive ability of order book imbalance for BTC prices on two exchanges, Bybit and OKEx, across different time scales, using the R score indicator.}
	\label{Orderbook Imbalance R Score}
\end{figure}

In our analysis, we explore the predictive ability of order book imbalance for BTC prices on two exchanges, Bybit and OKEx. The results are presented in Figure \ref{Orderbook Imbalance R Score}. Taking Bybit as an example, we use regression models for both a single exchange and cross-exchange signals to predict short-term price movements for BTC on Bybit. The regression models for the single exchange and cross-exchange signals are shown in equations \ref{single exchange} and \ref{cross exchange}, respectively. To simplify the analysis, we sum the order book imbalance of different exchanges to show their cumulative effect, rather than regressing the asset price on the order book imbalance of individual exchanges as separate variables.

\begin{equation}\label{single exchange}
    Ret_{t+1}^{i} = \alpha+\beta (IMB_{t}^{i})+\epsilon_{t}.
\end{equation}
\begin{equation}\label{cross exchange}
    Ret_{t+1}^{i} = \alpha+\beta (IMB_{t})+\epsilon_{t}.
\end{equation}

We evaluate the performance of the Order book Imbalance factor at different time intervals ranging from $100ms$ to $10s$, and compare the effects of using signals from a single exchange versus multiple exchanges. Figure A shows the experimental results on Bybit, while Figure B shows the results on OKEx. The results in Figure A indicate that the R score of the cross-exchange signal reaches 13\% at a time interval of 100ms, which is significantly higher than that of the single-exchange signal. Moreover, as the prediction time interval becomes longer, the predictive power of the signal decreases, as seen in the experiments on OKEx and Bybit. It should be noted that the cross-exchange signal can still improve the prediction level relative to the single-exchange signal, although the same factor has a weaker predictive effect on OKEx than on Bybit.

\begin{figure}
	\begin{center}
	\includegraphics[width=0.98\columnwidth]{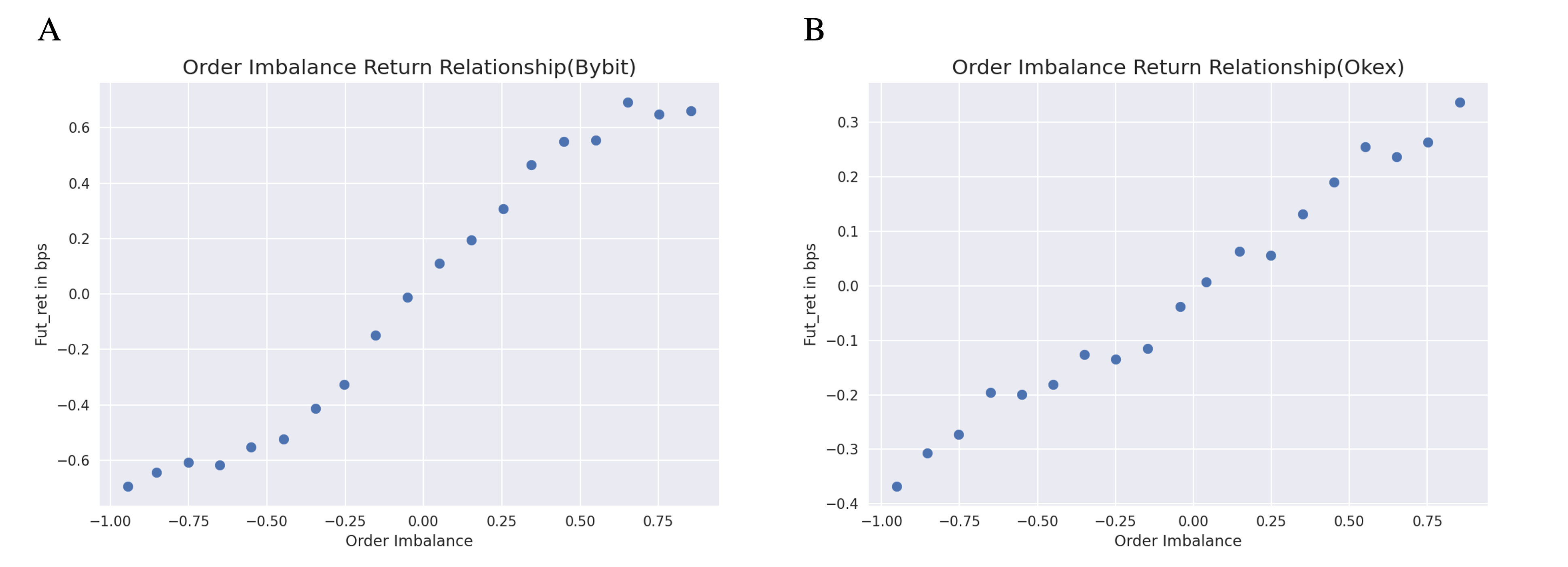}
	\end{center}
	\caption{\textbf{Future return prediction 5s (BTC\_USDT).} This figure displays the results of analyzing the level of future return predicted by the signal at 5-second intervals.}
	\label{Orderbook Imbalance future return}
\end{figure}

In addition to evaluating the predictive power of the Order book Imbalance factor, we also examine the level of the future return predicted by the signal at a time interval of 5s, as shown in Figure \ref{Orderbook Imbalance future return}. Figures A and B respectively present the experimental analysis on Bybit and OKEx. Taking Figure A as an example, we observe a positive correlation between the strength of the signal and the absolute size of the asset's future return. As the signal becomes stronger, the absolute magnitude of the future return also increases. When the signal tends to -1 or +1, the absolute value of the future return is roughly around 0.6bps, indicating that the maximum predictive power of the signal is around 0.6bps. The predictive performance of the signal on OKEx is not as significant as on Bybit, with an approximate prediction level of 0.4bps. 

\subsection{Cross Exchange Spread}
Due to the diversity of virtual currency trading venues, price differences between different exchanges are common. These differences are primarily caused by variations in market depth and trading populations across different exchanges. For instance, in Korean exchanges, trading volumes may be concentrated during the daytime in Asia due to the majority of traders being Koreans. The emergence of price differences has created a new type of market participant, the arbitrageurs, who aim to profit by buying low and selling high between different exchanges. These market participants help maintain asset prices at similar levels across different markets, bringing about a regression property in asset prices. As a result, cross-exchange price difference signals have become an important factor to consider in predicting short-term price changes of assets. In fact, Albers et al. \cite{albers2021fragmentation} conducted relevant research on predicting short-term price changes of assets based on cross-exchange spread signals.

Cross-exchange arbitrage is a strategy used to take advantage of price differences between different exchanges. The goal of this strategy is to profit by buying an asset on one exchange where it is undervalued and selling it on another exchange where it is overvalued. This arbitrage behavior helps to bring about a convergence in prices between the two exchanges over a short period of time. For example, when the price of an asset on exchange A is higher than that on exchange B, arbitrageurs will buy the corresponding assets on exchange B while selling the same amount of assets on exchange A. This buying behavior on exchange B increases the demand for the asset and raises its price, while the selling behavior on exchange A increases the supply of the asset and lowers its price. As a result, the prices of the asset on the two exchanges will gradually converge towards the same level.

In our analysis, we define the cross-exchange spread signal as shown in Equation \ref{cross exchange spread}. Here, $i$ and $j$ represent different exchanges, and when we consider exchange $i$, the spread between exchange $i$ and exchange $j$ can be used as a factor to predict the price change of exchange $i$. If $Spread_{t}^i>0$, it indicates that prices on other exchanges are higher than that on exchange $i$, while $Spread_{t}^i<0$ indicates that prices on other exchanges are lower than that on exchange $i$. In addition, we consider the potential impact of derivatives on prices, where factors such as funding rates may lead to a relatively persistent spread between derivatives and spot prices. Therefore, we also consider a normalized version of the spread by subtracting a rolling mean from the original spread, as shown in Equation \ref{normalized cross exchange spread}. This normalization helps to account for any persistent spread caused by derivatives trading and allows for a more accurate analysis of price movements based on cross-exchange spread signals.

\begin{equation}\label{cross exchange spread}
    Spread_{t}^{i} = \sum_{j}(Price_{t}^{j}-Price_{t}^{i}).
\end{equation}
\begin{equation}\label{normalized cross exchange spread}
    Spread\_norm_{t}^{i} = Spread_{t}^{i}-Roll(Spread_{t}^{i}).
\end{equation}

\begin{figure}
	\begin{center}
	\includegraphics[width=0.98\columnwidth]{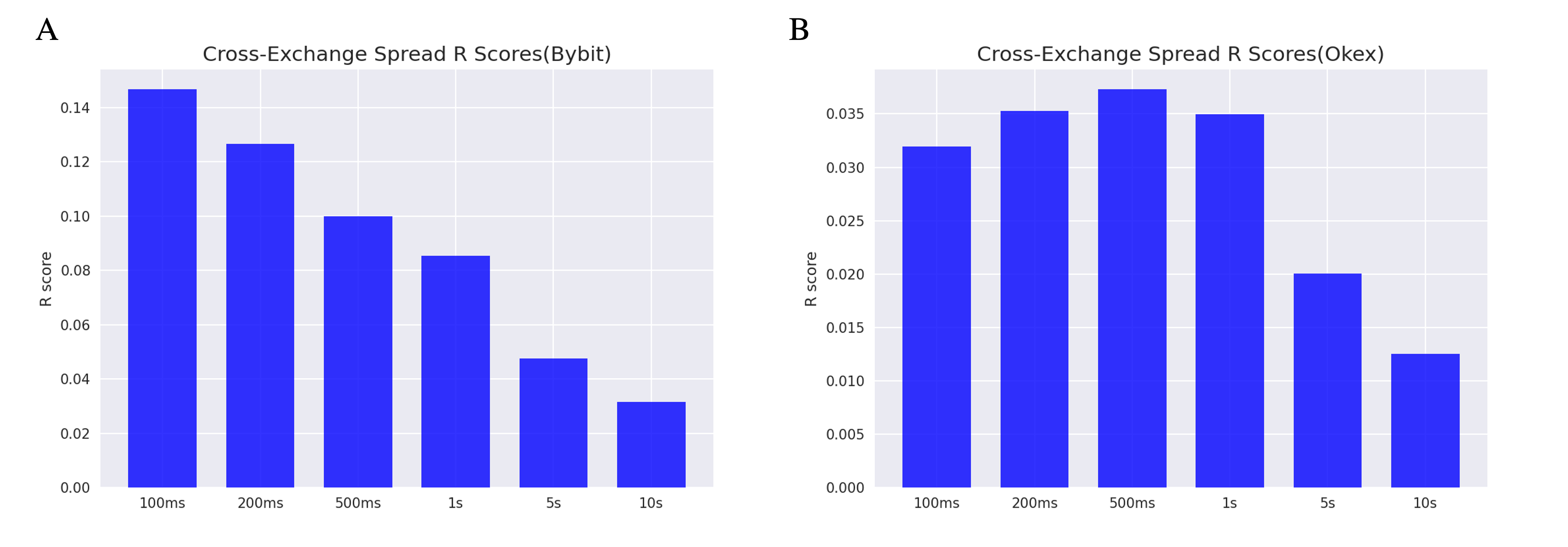}
	\end{center}
	\caption{\textbf{Cross Exchange Spread R Score (BTC\_USDT).} The regression analysis results for the signal cross-exchange spread in Bybit and OKEx.}
	\label{Cross Exchange Spread R score}
\end{figure}

The regression analysis results are shown in Figures \ref{Cross Exchange Spread R score} and \ref{Cross Exchange Spread R score} for Bybit and OKEx exchanges, respectively. Although the explanatory power of the factor decreases as the prediction time interval becomes larger, the decay rate is not very fast. This is also reflected in the experiments on OKEx, where we found that the explanatory power of the factor even reached its maximum at a time interval of 500ms and did not always decay. Moreover, the experimental results show that compared to the order flow signal, the cross-exchange spread has a special advantage in longer time intervals. This may be because the convergence of cross-exchange spread regression does not occur instantly, but requires some time.

\begin{figure}
	\begin{center}
	\includegraphics[width=0.98\columnwidth]{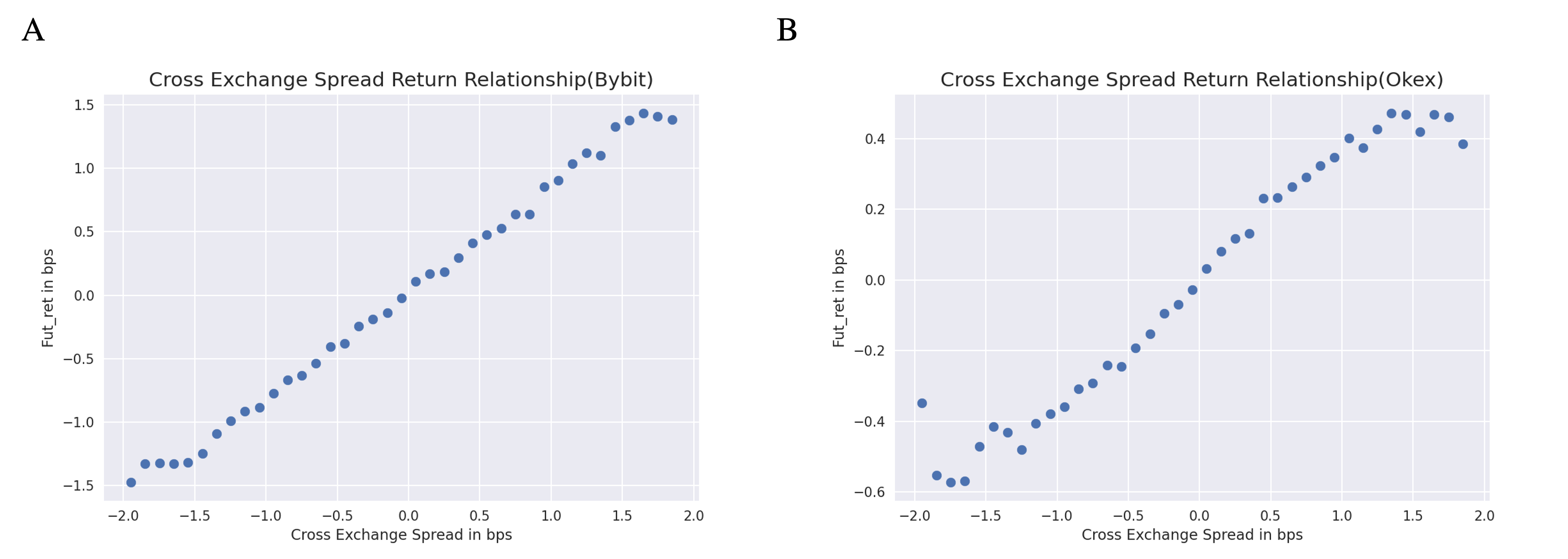}
	\end{center}
	\caption{\textbf{Future return prediction 5s (BTC\_USDT).} The relationship curve between the factor and future return.}
	\label{Cross Exchange future return}
\end{figure}

We can see from the relationship curve in Figure \ref{Cross Exchange future return} between the factor and future return that the predictive power of the factor is improved compared to the order flow signal. As the strength of the factor gradually increases, the future return tends to 1.5bps, indicating that the maximum predictive power of the factor tends to 1.5bps. Furthermore, there is a positive correlation between the strength of the factor and future return, which can be helpful in the decision-making process.

\begin{figure}
	\begin{center}
	\includegraphics[width=0.98\columnwidth]{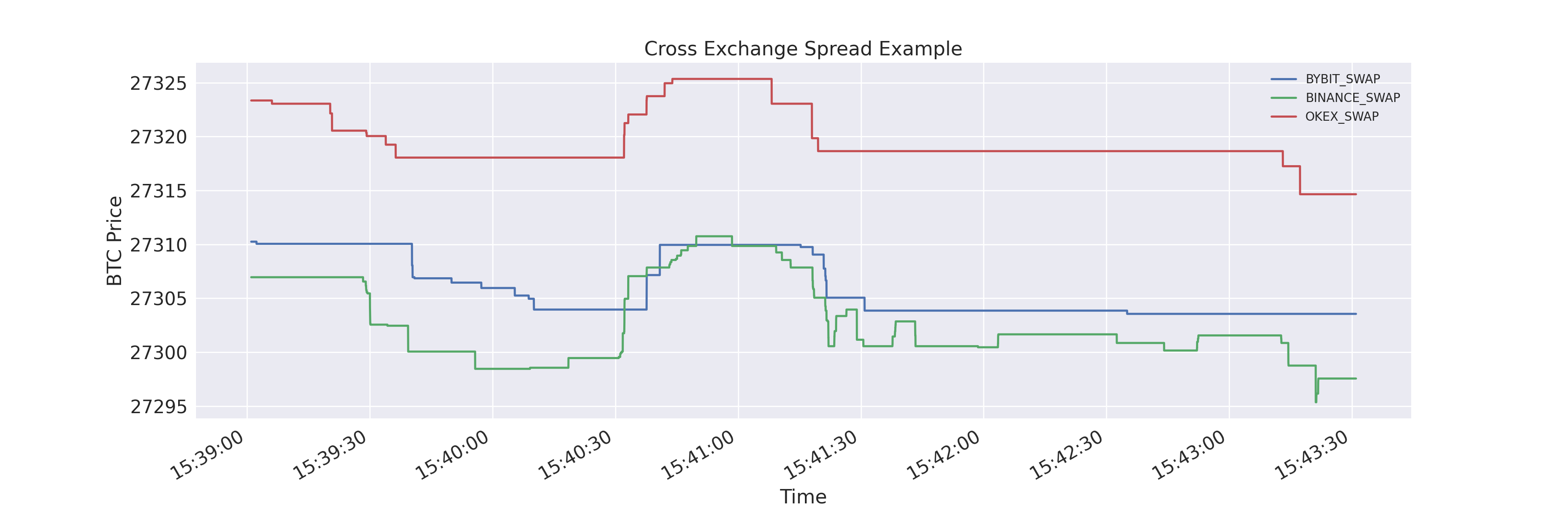}
	\end{center}
	\caption{\textbf{Cross Exchange Spread Example (BTC\_USDT).} The figure shows the price curves of BTC in three exchanges during the same time period. We can observe that price changes first occur on Binance and OKEX, with Bybit's price changes lagging behind the other two.}
	\label{Cross Exchange Spread Example}
\end{figure}

To further illustrate the potential reasons why cross-exchange spread signals can predict asset prices, we provide a simple example in Figure \ref{Cross Exchange Spread Example}. The figure shows the price curves of BTC in three exchanges during the same time period. Several phenomena are evident: Firstly, there is a relatively stable price difference between exchanges, with the price of OKEX being slightly higher than the other two exchanges. If we focus on the sequence of BTC price changes, we can observe that price changes first occur on Binance and OKEX, with Bybit's price changes lagging behind the other two. When BTC rises on Binance, a cross-exchange spread is created between Bybit and Binance, which typically leads to an increase in the price of Bybit. This is a straightforward example of using cross-exchange spread signals to predict asset prices.

\section{Optimal Execution Using RL}
When executing a large order, the market price can be significantly affected, especially when the order size is comparable to or larger than the market liquidity. If the order is a large buy order, the market price will rise, whereas if it is a sell order, the market price will fall. In order to mitigate the impact on the market price, we need to split the large order into multiple sub-orders during the execution process. Executing orders too quickly can adversely affect the market price, while executing orders too slowly can expose us to too much market risk (price uncertainty). Therefore, we need to make appropriate decisions based on the market conditions at the time.

\subsection{Problem Formulation}

Consider selling V units of stocks within a given time T, where $Q_{t}$ represents the current inventory, $\nu_t$ represents the execution speed, and $S_{t}$ represents the price of the stock at time t. The current inventory depends on the execution speed and can be modeled using the function \ref{inventory function}. 
\begin{equation}\label{inventory function}
    \begin{split}
        dQ_{t} &= -\nu_{t}dt, \\
        Q_{0} &= V.
    \end{split}
\end{equation}
To account for the impact of execution speed on the execution price, we consider the execution price $\hat{S}{t}$ as a function of the current stock price $S{t}$ and the execution speed $\nu_{t}$, which can be represented by the following equation \ref{execution price}. Here, the variable k characterizes the extent to which execution speed affects the execution price.
\begin{equation}\label{execution price}
    \hat{S}_{t} = S_{t}-k\nu_{t}.
\end{equation}
Our goal is to find the optimal execution speed $\nu_t$ that enables us to execute the given number of stocks as quickly as possible, without adversely affecting the execution price and while achieving the highest average execution price. In other words, we aim to maximize the following objective function \ref{objective function}:
\begin{equation}\label{objective function}
    Q = \int_0^T \hat{S}_u \nu_u d u+\left(V-Q_T\right) S_T-\alpha\left(V-Q_T\right)^2,
\end{equation}
the first term in the objective function represents the cash obtained from selling stocks accumulated within the time interval from 0 to T. The integral term in the objective function represents the sum of the average execution price of selling stocks at time $u$ times the number of stocks sold at time $u$, represented by $\hat{S}_u \nu_u du$. The second term represents the stocks we still hold but have not sold until time T, and our strategy is to sell these stocks at the stock price $S_T$ at time T. The third term is a penalty term, which ensures that the execution algorithm can sell all stocks before time T, an important prerequisite for optimal execution.

When making decisions about execution speed, we need to consider both the current position level and the remaining execution time to ensure that we can sell all stocks within the specified time. Optimal execution also focuses on the execution price, meaning that we aim to sell stocks at a higher price. Therefore, when making decisions about execution speed, we also need to pay attention to the market situation and predict the short-term trend of stock prices. For instance, if we anticipate that the price will rise, we can slow down the execution speed and sell most of the stocks after the rise. Conversely, if the price is expected to fall, we can speed up the execution speed to ensure the stocks are fully executed before the price drops.

\subsection{RL Environment For Optimal Execution}
\subsubsection{States}

The state S is divided into two parts. One part is generated by the agent itself and is called the agent state, denoted by $S_{agent}$. These variables mainly originate from the task itself, which involves selling V units of stocks within T time. There are two crucial variables for our decision-making: the current inventory level $Q_{t}$ and the remaining execution time $M=T-t$. To simplify the optimal execution decision-making process, we discretize the problem and introduce the concept of decision resolution H. For example, if the execution time is T=10s, the number of stocks to be executed is V=1000, and the decision resolution is H=10, it means that we can make a decision every 1 second on the number of stocks to be executed, and the remaining time ranges from 10 (at the beginning of the task) to 0 (at the end of the task).

Another type of variable is the environmental variable, which originates from our observation of the environment. We refer to it as the environment state, denoted by $S_{environment}$. In the context of the optimal execution problem, it mainly involves predicting factors related to the future return of the asset. Much of the work focuses on analyzing the current state of the market, including extracting factors related to the current market order flow and order book shape. Our work primarily focuses on the analysis and utilization of cross-exchange signals, such as cross-exchange order book imbalance, cross-exchange trade flow imbalance, and cross-exchange spread analysis.

Our focus is on analyzing the impact of different states on the decision-making of the agent, which includes the current inventory, remaining time, and various market variables.

\subsubsection{Actions}

In the reinforcement learning model, the agent typically makes decisions, or actions, based on the current state. In the context of the optimal execution problem, we consider the classic TWAP (Time-Weighted Average Price) model as the baseline for model evaluation. For selling V units within a given time T, the TWAP algorithm provides a solution that executes the order at a constant speed of $\nu = V/T$. In the discrete case, given a decision resolution of H, the TWAP execution algorithm needs to sell V/H units of orders every T/H time interval, with the orders placed in the form of market orders.

To maintain consistency with the TWAP execution algorithm, we consider market orders as the order type, with the only variable being the number of stocks to be executed at different times. Unlike the uniform execution of the TWAP algorithm, we allow the number of market orders to be any integer in the interval $[0, Q_t]$, where 0 represents no execution and $Q_t$ represents the current position level at time t. When the quantity of market orders is equal to the current position level, it represents selling all orders at that moment, which typically occurs when the remaining time is close to 0 or when the prediction signal is extremely bearish.

\subsubsection{Rewards}

When the agent makes a decision $a_t$, the environment provides a corresponding reward. The design of the reward function is typically the most important aspect of reinforcement learning algorithms. In this case, we use the relative P\&L (Profit and Loss) between time $t$ and $t+1$ as the reward function, as shown in the following equation:
\begin{equation}\label{reward function}
    R_t = \frac{Q_{t+1}*(S_{t+1}-S_{t})-\beta|(Q_{t+1}-Q_{t})*S_{t}|}{V*S_0}.
\end{equation}
The variable $Q_{t+1}$ in the formula represents the current position level. At time $t$, we sell $(Q_{t}-Q_{t+1})$ shares of the stock, where the constant $\beta$ represents the transaction cost, including transaction fees and impact costs of selling stocks. To avoid introducing bias caused by different stock prices, we consider the relative return with respect to the initial trading price. $V$ represents the total number of shares we need to sell, $S_0$ is the price of the stock at time 0, and $S_0*V$ is the total cash we would obtain if we sold all shares instantaneously at time 0, assuming infinite liquidity.

In contrast to previous approaches that only consider the realized implementation shortfall \cite{schnaubelt2022deep} and absolute proceeds \cite{nevmyvaka2006reinforcement}, we believe that it is essential to consider the impact of current actions on future outcomes. Additionally, this reward design has the advantage that the accumulated reward over an episode is a commonly used performance metric for optimal execution, known as implementation shortfall.

\subsubsection{Reinforcement Learning Algorithm}

We adopted Proximal Policy Optimization (PPO) as our reinforcement learning algorithm model. PPO is a variant of the Actor-Critic model, consisting of two main components: the actor, which outputs the probability distribution of actions given the current state $S_t$, and the critic, which estimates the state value function. The actor takes the state $S_t$ as input and outputs a stochastic probability distribution over $a_t$. The critic is used to estimate the state value function and from it we obtain the advantage function. By maximizing the advantage function, we can optimize the actor's action space.

The PPO algorithm (Proximal Policy Optimization) originated from the TRPO algorithm (Trust Region Policy Optimization)\cite{schulman2015trust}. By constraining the rate of gradient updates, PPO significantly improves the stability and convergence speed issues in policy optimization. The objective function of PPO algorithm consists of three main parts \ref{ppo loss function}:
\begin{equation}\label{ppo loss function}
    L_t^{C L I P+V F+S}(\theta)=\hat{\mathbb{E}}_t\left[L_t^{C L I P}(\theta)-c_1 L_t^{V F}(\theta)+c_2 S\left[\pi_\theta\right]\left(s_t\right)\right].
\end{equation}
\begin{equation}\label{ppo clip function}
    L^{C L I P}(\theta)=\hat{\mathbb{E}}_t\left[\min \left(r_t(\theta) \hat{A}_t, \operatorname{clip}\left(r_t(\theta), 1-\epsilon, 1+\epsilon\right) \hat{A}_t\right)\right].
\end{equation}

The first component of the objective function \ref{ppo clip function} is mainly designed to limit the update speed of the policy, where the hyperparameter $\epsilon$ is a constant and the ratio $r_t(\theta)=\frac{\pi_{\theta}(a_t|s_t)}{\pi_{\theta_{old}}(a_t|s_t)}$ is used to measure the distance between the policy $\theta$ and $\theta_{old}$. The second component is the squared error loss of the value function $L_t^{VF}=(V_{\theta}(s_t)-V_{t}^{target})^2$, which is used to update the parameters of the critic model. The third component is an entropy loss based on the current policy $\pi_{\theta}$, which is mainly designed to prevent overfitting of the model in the early stages.

\section{Results}
Before presenting our experimental results, let's briefly review the entire experimental process. We used high-frequency data of cryptocurrencies as our experimental object. Considering the diversity of cryptocurrency trading venues, we are more focused on cross-exchange signals, rather than just collecting signals through single exchange order flow. We use high-frequency data to extract features and input them as market state variables into the reinforcement learning model. Based on the market state, the reinforcement learning model provides optimal execution decisions and updates model parameters through market feedback rewards.
\subsection{Environment and Model Setting}
We consider selling a virtual currency of $V=50$ units within a specified execution time of $T=50$ seconds. To simplify the optimal execution problem, we consider a discrete case where the decision resolution is $H=10$, meaning that the agent can make 10 decisions within the given time, i.e., every 5 seconds. In the discrete case, the TWAP execution algorithm will sell 5 units of virtual currency evenly at each decision time. The goal of reinforcement learning is to better allocate the order execution quantity at each time point so that the average selling price is higher than that of the TWAP execution algorithm.

For the PPO policy, we used a two-layer fully connected neural network with 64 neurons in each layer. For the PPO algorithm hyperparameters, we set $\epsilon=0.2$, $\gamma=0.99$, and $\lambda=0.95$. Here, $\epsilon$ is the importance sampling ratio used to control the difference between the new and old policies, $\gamma$ is the discount factor used to weigh short term and long term rewards, and $\lambda$ is the weighting factor of the advantage function. We used Adam optimizer with a policy learning rate of 0.0003 and a value function estimation network learning rate of 0.001.

\subsection{Evaluation Metrics}

To evaluate the performance of the optimal execution algorithm, we consider the total cash obtained by selling all assets, denoted as Cash in \ref{cash}.

\begin{equation}\label{cash}
    Cash = \sum_t^T(x_t*p_t)+(V-Q_T)*S_T-\alpha*((V-Q_T)^2)*S_T.
\end{equation}

Here, $x_t$ represents the number of stocks sold at time t, $p_t$ represents the selling price of the stock at time t, $V-Q_T$ represents the outstanding order quantity that has not been executed at time T, and $\sum_t^T(x_t*p_t)+(V-Q_T)S_T$ represents all the cash obtained from selling the assets. The last term is a penalty term used to penalize the execution algorithm for not completing all the orders that need to be executed within a fixed time, with a value of $\alpha=0.02$. In addition to the zero ending penalty, we also compare the performance of strategies considering price impact in the subsequent experiments.

Noticing that the total amount of cash is an absolute value, we consider using implementation shortfall as a relative indicator, as shown in equation \ref{implementation shortfall}:

Where $V*p_0$ represents the cash obtained by selling all orders at time 0. Implementation shortfall is a common measure in the field of optimal execution, which evaluates the performance of an execution algorithm by comparing the average execution price with the initial price $p_0$.

\begin{equation}\label{implementation shortfall}
    IS = (Cash-V*p_0)/(V*p_0).
\end{equation}

We focus on the relative performance of the execution algorithm compared to TWAP and define the variable Gain in \ref{gain}
\begin{equation}\label{gain}
    Gain = (IS_{model}-IS_{twap})*10^4.
\end{equation}

\subsection{Statistical Results}
The experimental section mainly compares the statistical results of three different scenarios:
\begin{enumerate}
\item TWAP(Time weighted average price) execution
\item Consider only a single exchange signal as the input to the reinforcement learning model.
\item Consider cross exchange signal as the input to the reinforcement learning model.
\end{enumerate}

\begin{figure}
	\begin{center}
	\includegraphics[width=1\columnwidth]{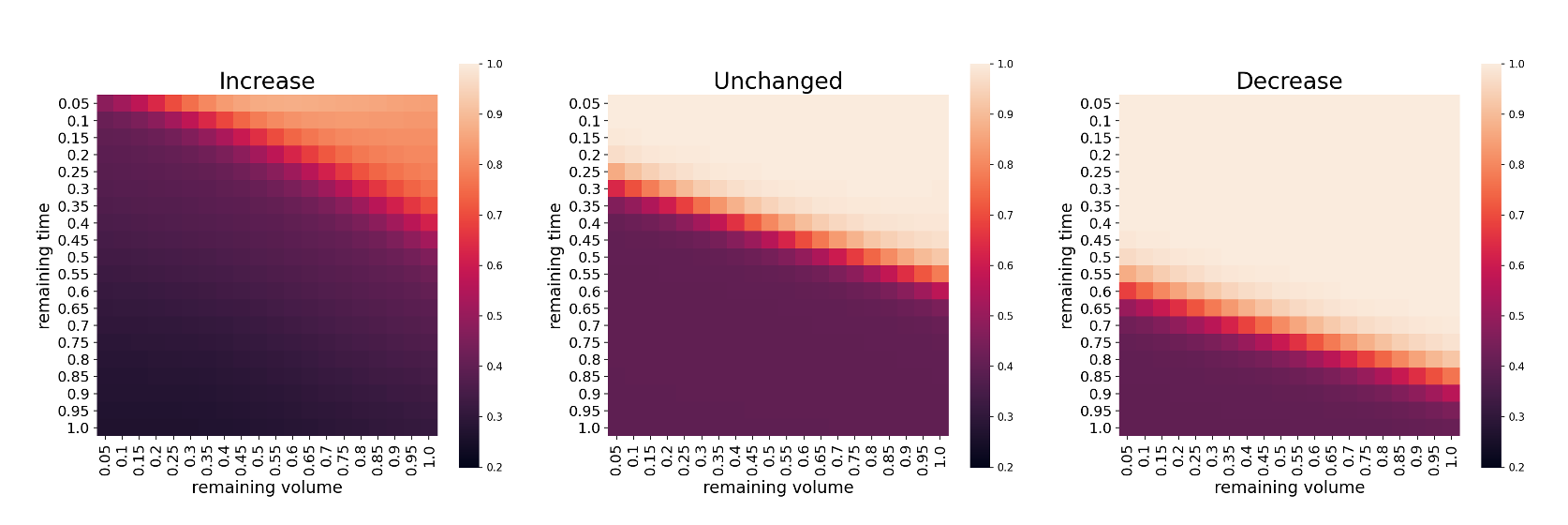}
	\end{center}
	\caption{\textbf{Action Space.} We categorize the market prediction signal into three categories: Increase, Unchanged, and Decrease. For each prediction signal, we visualize the action space given by the agent as a heatmap. It is apparent that the action space is a function of the remaining time and remaining volume. The color of the heatmap represents the aggressiveness of the execution, with lighter colors indicating more aggressive order execution.}
	\label{Action space}
\end{figure}

\subsubsection{Action Space}

To illustrate how the strategy is affected by state variables, we consider the relationship between state variables and the action space, as shown in Figure \ref{Action space}. The state variables include remaining time, inventory level, and market prediction signal. We normalize the remaining time and inventory level to the interval [0,1]. A remaining time of 1 represents the beginning of the episode, and a remaining time of 0 represents the end of the episode. A remaining volume of 1 represents a high inventory level, where no orders have been executed, and a remaining volume of 0 represents that all assets have been sold, and the order execution task has been completed.

For the market prediction signal, we divide it into three categories: Increase, Unchanged, and Decrease. Increase represents a prediction of an upward trend in the market, Decrease represents a prediction of a downward trend, and Unchanged represents a neutral signal with similar probabilities of an increase or decrease. Meanwhile, we map the aggressiveness of the orders executed by the agent to [0,1], where 1 represents the need for aggressive execution and 0 represents a passive execution.

We visualize the action space given by the agent in different states as a heatmap, as shown in Figure \ref{Action space}. From left to right, the action space in three situations are displayed: predicting an upward trend, predicting a neutral market, and predicting a downward trend. It can be observed that the action space is a function of remaining time and remaining volume. Taking the left heatmap as an example, it can be seen that as the remaining time decreases, the color of the heatmap becomes brighter, indicating that the order needs to be executed more aggressively. At the same time, as the position level increases, the aggressiveness of the order executed by the agent also increases. Regarding the market variables, by comparing the three heatmaps, it can be observed that when the prediction signal is a downward trend, the aggressiveness of the execution actions given by the agent is relatively high. This means that when the market is expected to decline, the agent usually wants to sell the position at a faster rate.

\begin{figure}
	\begin{center}
	\includegraphics[width=0.98\columnwidth]{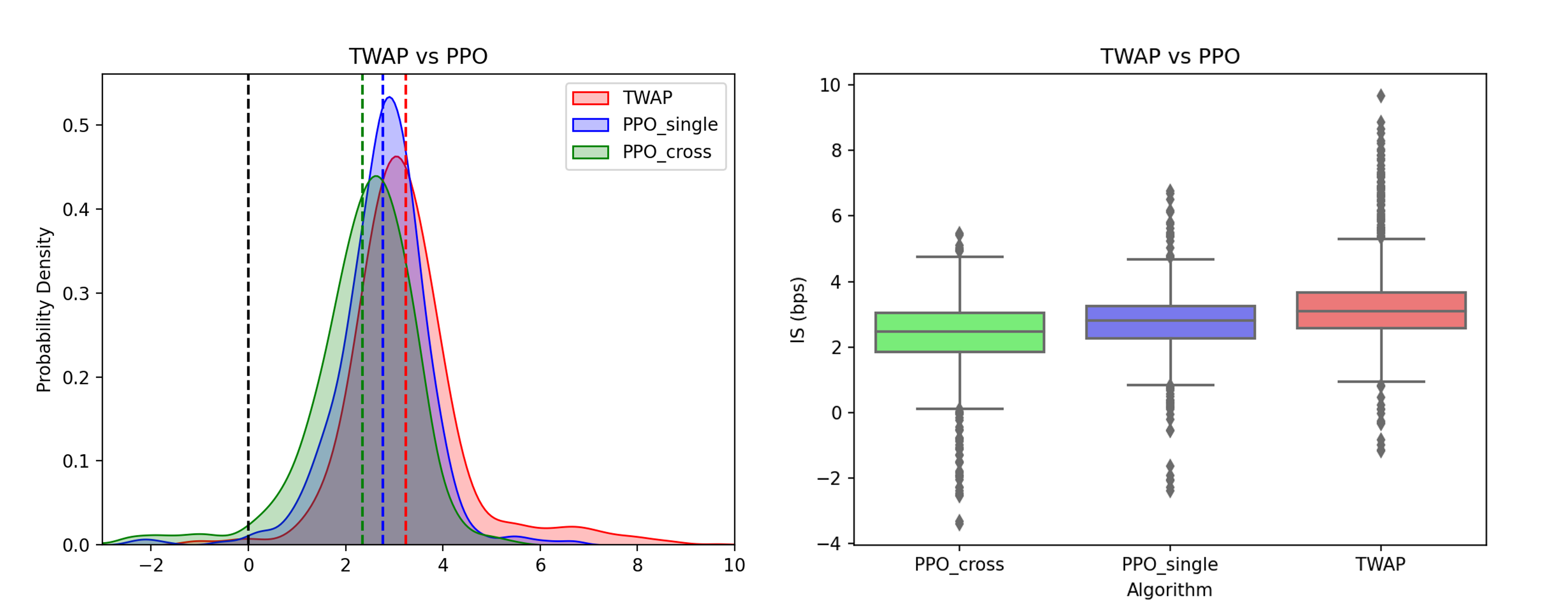}
	\end{center}
	\caption{\textbf{Algorithm comparison based on P\&L distribution (BTC\_USDT).} This figure shows the comparison results among three different algorithms. The test data consists of 1000 randomly sampled different samples from the test set, and the corresponding implementation shortfall is obtained by simulating the order execution process in real-world scenarios.}
	\label{Algorithm comparison}
\end{figure}

\begin{table}[ht]
\centering
\resizebox{0.8\textwidth}{!}{
\begin{tabular}{|c|c|c|c|}
\hline
 & $IS_{mean}$(bps) & $IS_{variance}$(bps) & $Gain_{twap}$(bps) \\
\hline
TWAP & -3.10 & 1.15 &  0\\
\hline
$PPO_{single}$ & -2.72 & 0.95 & 0.38 \\
\hline
$PPO_{cross}$ & -2.42 & 1.01 & 0.68 \\
\hline
\end{tabular}
}
\caption{\textbf{Algorithm Comparison(Bybit).} This table shows the comparison results among three different algorithms using Bybit data. The test data consists of 1000 randomly sampled different samples from the test set. }
\label{tab:comparison(Bybit)}
\end{table}

\begin{table}[ht]
\centering
\resizebox{0.8\textwidth}{!}{
\begin{tabular}{|c|c|c|c|}
\hline
 & $IS_{mean}$(bps) & $IS_{variance}$(bps) & $Gain_{twap}$(bps) \\
\hline
TWAP & -1.48 & 2.04 &  0\\
\hline
$PPO_{single}$ & -1.38 & 1.14 & 0.1 \\
\hline
$PPO_{cross}$ & -1.21 & 1.15 & 0.27 \\
\hline
\end{tabular}
}
\caption{\textbf{Algorithm Comparison(OKEx).} This table shows the comparison results among three different algorithms using OKEx data. The test data consists of 1000 randomly sampled different samples from the test set.}
\label{tab:comparison(OKEx)}
\end{table}

\subsubsection{Without Price Impact}

Table \ref{tab:comparison(Bybit)} and \ref{tab:comparison(OKEx)} shows the comparison results among three different algorithms. The test data consists of 1000 randomly sampled different samples from the test set, and the corresponding implementation shortfall is obtained by simulating the order execution process in real-world scenarios. The corresponding histogram is shown in \ref{Algorithm comparison}.

It can be seen that considering the execution fee of 3 bps, the IS of the TWAP execution algorithm is about -3.1 bps, which means that most of the execution costs are consumed by the execution fee. Due to the lack of price prediction, the P\&L due to price changes is roughly zero. The second algorithm uses a signal from a single exchange for price prediction to assist in the optimal execution process. Compared to the TWAP execution algorithm, the execution algorithm has improved by 0.38 bps, which means that the execution algorithm has gained some improvement by predicting asset price fluctuations. The third algorithm adds cross-exchange signals on the basis of single-exchange signals. Compared to the TWAP execution algorithm, the agent further gains 0.68 bps of improvement. This result further proves that cross-exchange signals can provide additional improvement compared to using only single-exchange signals.

We also noticed that the variance increases to some extent after using cross-exchange signals. We can understand this as follows: due to the different locations of servers of different exchanges, there may be network delay issues in the process of obtaining cross-exchange signals, which may lead to significant delays in the data collection process, especially in the case of large market conditions. The appearance of significant delays may cause inaccuracies in the signals obtained in some cases, which may be the main reason for the slightly increased variance in the experimental results.

\subsubsection{Price Impact}
So far, we have always assumed that our own orders have no impact on the market price. However, in real-world applications, large order sizes may have a significant impact on the market price. To test the robustness of the order execution strategy, we introduce the price impact function \ref{price impact function}. In the function, the variable $x_t$ represents the order size, V is the target execution volume, and $S_0$ is the initial price of the stock. For the TWAP execution algorithm, with a decision granularity of 10, the additional cost will be equal to 0 since the order size is equal to $0.1V$. As the order size $x_t$ becomes larger, the value of the price impact function also becomes larger, resulting in greater costs. In the experiment, we set the parameter $\beta$ to 1e-5. In order to ensure that the algorithm can complete the execution task within a specified time, we consider the zero-ending penalty \ref{zero ending penalty}, where the parameter $\alpha$ is a constant used to control the strength of the penalty. In the experiment, we set the parameter $\alpha$ to 0.02.

\begin{equation}\label{price impact function}
    C = \beta*max(0,x_t/V-0.1)*V*S_0.
\end{equation}
\begin{equation}\label{zero ending penalty}
    Z = \alpha*((V-Q_T)^2)*S_T.
\end{equation}

The data in Table \ref{tab:comparison_cost} shows the comparison results after considering the price impact function and zero ending penalty. As expected, the execution cost has increased significantly. Compared to the previous scenario where the price impact was not considered, the improvement obtained by the reinforcement learning model has slightly decayed. However, we can still see that compared to the TWAP execution algorithm, the reinforcement learning model can achieve lower execution costs. Moreover, compared to using signals from a single exchange, cross-exchange signals can still provide more predictive signals. After considering cross-exchange signals, the execution cost can be reduced by 0.32 bps compared to the TWAP execution algorithm.

\begin{table}[ht]
\centering
\resizebox{0.8\textwidth}{!}{
\begin{tabular}{|c|c|c|c|}
\hline
 & $IS_{mean}$(bps) & $IS_{variance}$(bps) & $Gain_{twap}$(bps) \\
\hline
TWAP & -3.10 & 1.15 &  0\\
\hline
$PPO_{single}$ & -2.96 & 0.78 & 0.14 \\
\hline
$PPO_{cross}$ & -2.78 & 0.66 & 0.32 \\
\hline
\end{tabular}
}
\caption{\textbf{Algorithm Comparison(Price impact).} This table shows the comparison results after considering the price impact function and zero ending penalty.}
\label{tab:comparison_cost}
\end{table}

\subsubsection{Influence Of Prediction Time Interval}
The above experiments mainly focused on predicting a time interval of 5 seconds. In order to investigate the impact of prediction time intervals on strategy performance, we compared the performance of three different time dimensions (2s, 5s, 10s). During the experiment, we kept the decision resolution H=10 unchanged. The experimental results are shown in Table \ref{tab: Prediction Time Interval}.

We mainly focus on the expectation and variance of the implementation shortfall metric, as well as the improvement compared to the TWAP baseline. From the perspective of expected execution cost, we observe that when the time interval is 5 seconds, the execution cost is the smallest at 2.42bps, followed by 2.64bps for the 10-second execution period, and the worst is the 2-second execution period. This can be explained from the absolute value of the expected return prediction: first, in the feature analysis chapter, we can see that as the prediction period becomes longer, the predictive power of the factor will decrease secondly, as the prediction period becomes longer, the amplitude of asset price fluctuations will increase. The first point leads to a decrease in execution cost as the prediction period becomes shorter, while the second point is exactly the opposite, with execution cost decreasing as the prediction period becomes longer. Looking at the variance of execution cost, we notice that as the execution period becomes longer, the variance also becomes larger. This is mainly due to the increasing uncertainty of asset prices as the execution period becomes longer.

\begin{table}[ht]
\centering
\resizebox{0.8\textwidth}{!}{
\begin{tabular}{|c|c|c|c|}
\hline
 & $IS_{mean}$(bps) & $IS_{variance}$(bps) & $Gain_{twap}$(bps) \\
\hline
2s & -2.73 & 0.54 &  0.29 \\
\hline
5s & -2.42 & 1.01 & 0.68 \\
\hline
10s & -2.64 & 1.21 & 0.52 \\
\hline
\end{tabular}
}
\caption{\textbf{Influence Of Prediction Time Interval($PPO_{cross}$).} This table compared the performance of three different time dimensions (2s, 5s, 10s).}
\label{tab: Prediction Time Interval}
\end{table}

\subsubsection{Examples}
\begin{figure}
	\begin{center}
	\includegraphics[width=0.72\columnwidth]{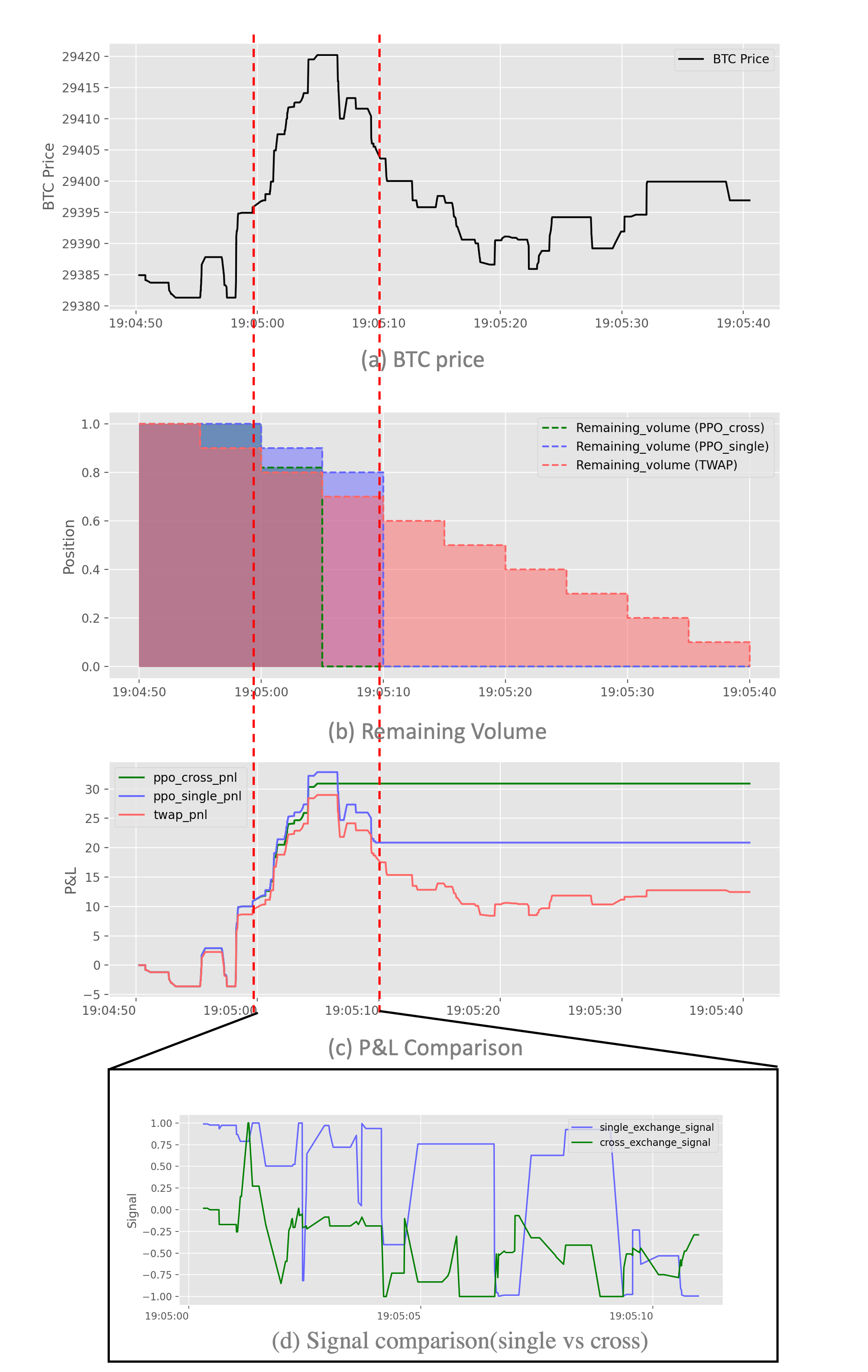}
	\end{center}
	\caption{\textbf{Execution Example.} This figure provides a simple example to compare three different execution algorithm. \textbf{(a)} This chart shows the price curve of Bitcoin (BTC) during the execution period. \textbf{(b)} This figure reflects the execution process of three algorithms. \textbf{(c)} This figure is a comparison of the P\&L (profit and loss) generated from the execution process. \textbf{(d)} This chart reflects the specific comparison between the cross-exchange signal and the single-exchange signal during the time interval from 19:05:00 to 19:05:10.}
	\label{Execution Example}
\end{figure}

To further explain the difference between the reinforcement learning execution algorithm and TWAP, we provide a simple example as shown in Figure \ref{Execution Example}. The execution starts at 12:47:40 and needs to execute 50 units of assets within 50 seconds. The solid blue line represents the price change of assets during this period. It can be seen that there was a significant downward trend in asset prices after the order execution began. The blue and yellow dotted lines in the figure represent the execution degree of the execution algorithm. For the TWAP execution algorithm, as there is no additional signal provided by the market, the position level steadily decreases, representing that the algorithm sells assets at a uniform rate. For the reinforcement learning agent, it can be seen that the position level quickly drops to 0, indicating that the agent sold all assets in the initial short period of time, effectively avoiding losses caused by the downward trend in asset prices.

\section{Conclusion}
In this work, we demonstrate that reinforcement learning models can be used to address the problem of optimal order execution. Optimal order execution is crucial for both professional teams and individuals, especially for strategies with high turnover. If execution can be improved, then the strategy can achieve significant gains.

Our work focuses primarily on the following points: First, we built an order matching system to simulate the execution of orders on an exchange, as the interaction environment for reinforcement learning. Second, considering the problem of multiple virtual currency exchanges, we attempted for the first time to use cross-exchange signals to predict asset price changes. Although we need to consider the significant delay in obtaining cross-exchange market data, experimental results show that the signals still have a significant predictive effect. Finally, we observed that after adding prediction signals, especially cross-exchange signals, the agent can make better decisions.

\bibliographystyle{unsrtnat}
\bibliography{references}  

\begin{thebibliography}{18}
\providecommand{\natexlab}[1]{#1}
\providecommand{\url}[1]{\texttt{#1}}
\expandafter\ifx\csname urlstyle\endcsname\relax
  \providecommand{\doi}[1]{doi: #1}\else
  \providecommand{\doi}{doi: \begingroup \urlstyle{rm}\Url}\fi

\bibitem[Bertsimas and Lo(1998)]{bertsimas1998optimal}
Dimitris Bertsimas and Andrew~W Lo.
\newblock Optimal control of execution costs.
\newblock \emph{Journal of financial markets}, 1\penalty0 (1):\penalty0 1--50,
  1998.

\bibitem[Edelen et~al.(2013)Edelen, Evans, and Kadlec]{edelen2013shedding}
Roger Edelen, Richard Evans, and Gregory Kadlec.
\newblock Shedding light on “invisible” costs: Trading costs and mutual
  fund performance.
\newblock \emph{Financial Analysts Journal}, 69\penalty0 (1):\penalty0 33--44,
  2013.

\bibitem[Almgren and Chriss(2001)]{almgren2001optimal}
Robert Almgren and Neil Chriss.
\newblock Optimal execution of portfolio transactions.
\newblock \emph{Journal of Risk}, 3:\penalty0 5--40, 2001.

\bibitem[Huberman and Stanzl(2005)]{huberman2005optimal}
Gur Huberman and Werner Stanzl.
\newblock Optimal liquidity trading.
\newblock \emph{Review of finance}, 9\penalty0 (2):\penalty0 165--200, 2005.

\bibitem[Nevmyvaka et~al.(2006)Nevmyvaka, Feng, and
  Kearns]{nevmyvaka2006reinforcement}
Yuriy Nevmyvaka, Yi~Feng, and Michael Kearns.
\newblock Reinforcement learning for optimized trade execution.
\newblock In \emph{Proceedings of the 23rd international conference on Machine
  learning}, pages 673--680, 2006.

\bibitem[Dahl{\'e}n and Rantil(2018)]{dahlen2018optimized}
Olle Dahl{\'e}n and Axel Rantil.
\newblock Optimized trade execution with reinforcement learning, 2018.

\bibitem[Schnaubelt(2022)]{schnaubelt2022deep}
Matthias Schnaubelt.
\newblock Deep reinforcement learning for the optimal placement of
  cryptocurrency limit orders.
\newblock \emph{European Journal of Operational Research}, 296\penalty0
  (3):\penalty0 993--1006, 2022.

\bibitem[Ning et~al.(2021)Ning, Lin, and Jaimungal]{ning2021double}
Brian Ning, Franco Ho~Ting Lin, and Sebastian Jaimungal.
\newblock Double deep q-learning for optimal execution.
\newblock \emph{Applied Mathematical Finance}, 28\penalty0 (4):\penalty0
  361--380, 2021.

\bibitem[Bao and Liu(2019)]{bao2019multi}
Wenhang Bao and Xiao-yang Liu.
\newblock Multi-agent deep reinforcement learning for liquidation strategy
  analysis.
\newblock \emph{arXiv preprint arXiv:1906.11046}, 2019.

\bibitem[Cont et~al.(2021)Cont, Cucuringu, and Zhang]{cont2021price}
Rama Cont, Mihai Cucuringu, and Chao Zhang.
\newblock Price impact of order flow imbalance: Multi-level, cross-sectional
  and forecasting.
\newblock \emph{arXiv preprint arXiv:2112.13213}, 2021.

\bibitem[Cartea and Jaimungal(2016)]{cartea2016incorporating}
Alvaro Cartea and Sebastian Jaimungal.
\newblock Incorporating order-flow into optimal execution.
\newblock \emph{Mathematics and Financial Economics}, 10:\penalty0 339--364,
  2016.

\bibitem[Cont et~al.(2014)Cont, Kukanov, and Stoikov]{cont2014price}
Rama Cont, Arseniy Kukanov, and Sasha Stoikov.
\newblock The price impact of order book events.
\newblock \emph{Journal of financial econometrics}, 12\penalty0 (1):\penalty0
  47--88, 2014.

\bibitem[Bouchaud et~al.(2009)Bouchaud, Farmer, and Lillo]{bouchaud2009markets}
Jean-Philippe Bouchaud, J~Doyne Farmer, and Fabrizio Lillo.
\newblock How markets slowly digest changes in supply and demand.
\newblock In \emph{Handbook of financial markets: dynamics and evolution},
  pages 57--160. Elsevier, 2009.

\bibitem[Byrd et~al.(2020)Byrd, Palaparthi, Hybinette, and
  Balch]{byrd2020importance}
David Byrd, Sruthi Palaparthi, Maria Hybinette, and Tucker~Hybinette Balch.
\newblock The importance of low latency to order book imbalance trading
  strategies.
\newblock \emph{arXiv preprint arXiv:2006.08682}, 2020.

\bibitem[Yagi et~al.(2023)Yagi, Hoshino, Mizuta, et~al.]{yagi2023impact}
Isao Yagi, Mahiro Hoshino, Takanobu Mizuta, et~al.
\newblock Impact of high-frequency trading with an order book imbalance
  strategy on agent-based stock markets.
\newblock \emph{Complexity}, 2023, 2023.

\bibitem[Cartea et~al.(2018)Cartea, Donnelly, and
  Jaimungal]{cartea2018enhancing}
Alvaro Cartea, Ryan Donnelly, and Sebastian Jaimungal.
\newblock Enhancing trading strategies with order book signals.
\newblock \emph{Applied Mathematical Finance}, 25\penalty0 (1):\penalty0 1--35,
  2018.

\bibitem[Albers et~al.(2021)Albers, Cucuringu, Howison, and
  Shestopaloff]{albers2021fragmentation}
Jakob Albers, Mihai Cucuringu, Sam Howison, and Alexander~Y Shestopaloff.
\newblock Fragmentation, price formation and cross-impact in bitcoin markets.
\newblock \emph{Applied Mathematical Finance}, 28\penalty0 (5):\penalty0
  395--448, 2021.

\bibitem[Schulman et~al.(2015)Schulman, Levine, Abbeel, Jordan, and
  Moritz]{schulman2015trust}
John Schulman, Sergey Levine, Pieter Abbeel, Michael Jordan, and Philipp
  Moritz.
\newblock Trust region policy optimization.
\newblock In \emph{International conference on machine learning}, pages
  1889--1897. PMLR, 2015.

\end{thebibliography}






\end{document}